\documentclass[iop,openany]{emulateapj_15}  

\usepackage{graphics}
\usepackage{graphicx}
\usepackage{amssymb}
\usepackage{amsmath}
\usepackage{amstext}  
\usepackage{amsfonts}    
\usepackage{floatrow}
\usepackage{relsize}	
\usepackage{booktabs}

\usepackage{natbib}

\usepackage{subfigure}
\usepackage{float}  		

\usepackage{mathtools}
\newcommand\invisiblesection[1]{%
  \refstepcounter{section}%
  \addcontentsline{toc}{section}{\protect\numberline{\thesection}#1}%
  \sectionmark{#1}}

\usepackage{epstopdf} 						
\usepackage{epsfig}

\usepackage[usenames,dvipsnames]{color}
\usepackage[
colorlinks=true,	
linkcolor=WildStrawberry ,  
urlcolor=blue ,   	   
citecolor=RoyalBlue ,  
filecolor=blue, anchorcolor=blue, runcolor=blue, hyperfootnotes=true
]{hyperref}			    
\usepackage{hypernat} 												
\usepackage{etex} 													
\usepackage{bookmark} 												

\def\farcs{\hbox{$.\!\!^{\prime\prime}$}}
\def\fs{\hbox{$.\!\!^{\rm s}$}}

\renewcommand{\thempfootnote}{\arabic{mpfootnote}}
\shortauthors{\small{Zanardo et al.}} 
\shorttitle{\small{Detection of Linear Polarization  in the Radio Remnant of  Supernova 1987A}}

\begin{document}
\title{Detection of Linear Polarization in the Radio Remnant of Supernova 1987A}

\author{
Giovanna Zanardo\altaffilmark{1}, 
Lister Staveley-Smith\altaffilmark{1}, 
B. M. Gaensler\altaffilmark{2}, 
Remy Indebetouw\altaffilmark{3,4}, \\
C. -Y. Ng\altaffilmark{5},
Mikako Matsuura\altaffilmark{6,7},
\& A. K. Tzioumis\altaffilmark{8}
}

\affil{
 \altaffilmark{1}International Centre for Radio Astronomy Research, M468, University of Western Australia, \\Crawley, WA 6009, Australia; \href{mailto:giovanna.zanardo@gmail.com}{giovanna.zanardo@gmail.com}\\
 \altaffilmark{2}Dunlap Institute for Astronomy \& Astrophysics, University of Toronto, Toronto, ON M5S 3H4, Canada\\ 
 \altaffilmark{3}Department of Astronomy, University of Virginia, P.O. Box 400325, Charlottesville, VA 22904-4325, USA\\
 \altaffilmark{4}National Radio Astronomy Observatory, 520 Edgemont Rd, Charlottesville, VA 22903, USA\\
 \altaffilmark{5}Department of Physics, University of Hong Kong, Pokfulam Road, Hong Kong\\
 \altaffilmark{6}Department of Physics and Astronomy, University College London, Gower St., London WC1E 6BT, UK\\
 \altaffilmark{7}School of Physics and Astronomy, Cardiff University, QueenÕs Buildings, The Parade, Cardiff, CF24 3AA, UK\\ 
 \altaffilmark{8}CSIRO Astronomy and Space Science, Australia Telescope National Facility, PO Box 76, Epping, NSW 1710, Australia\\
}

\vspace{6.0mm}
\begin{abstract}
Supernova 1987A in the Large Magellanic Cloud 
has proven a unique laboratory to investigate particle acceleration in young supernova remnants. 
Here we report the first detection of linear polarization of the supernova's synchrotron emission from imaging observations at frequencies spanning 
from 20 
to 50 GHz, carried out with the Australia Telescope Compact Array between October 2015 and May 2016.  
The direction of the radio polarization,
corrected for Faraday rotation,
points to a primarily
radial magnetic field across the inner ring, encompassing both the reverse and forward shocks.
The magnetic field strength peaks over the high-emissivity eastern sites, where
efficient cosmic ray acceleration likely takes place 
under 
quasi-parallel shocks at high Mach numbers. 
The mean fraction of polarized emission in the brightest sites  is $2.7\pm0.2\%$  at  22 GHz and $3.5\pm0.7\%$ at 44 GHz. 
In the inner remnant,  
non-radial components of the polarized emission appear to be more prevalent. However, the low significance 
detection in the central regions limits interpretation.

\end{abstract}
\vspace{7mm}

\section{Introduction}
\label{Intro}
Supernova remnants (SNRs) are powerful particle accelerators. 
As a supernova (SN) blast wave propagates through the circumstellar medium (CSM),
electrons and protons 
trapped between upstream and downstream magnetic mirrors
gain energy 
via multiple traversals of the shock front \citep{dru83,kir96}. 
The accelerated particles 
generate further 
magnetic field fluctuations and local amplification \citep{bel04}, 
thus leading to increased acceleration efficiency \citep{koy95,fer13, fer10}. 
The geometry and orientation of the magnetic field that drive an efficient particle acceleration process by the shock front remain under debate. 

Although older SNRs have been observed with a preferentially tangential magnetic field, 
many young SNRs 
exhibit some degree of radial alignment \citep{rey93,rey12}.
Theoretical models show that the magnetic field lines can be stretched radially by the Rayleigh-Taylor (R-T) instability~\citep{gul73} at the contact discontinuity between the supernova ejecta and the compressed CSM. 
However, it is still unclear whether 
the R-T instability or another mechanism can reproduce a radial 
field that extends outward to the forward shock~\citep{bad16}.
\hspace{-0mm}
According to ideal magneto-hydrodynamics (MHD), 
a radial field that reaches 
the forward shock can be obtained 
when cosmic rays 
significantly contribute to 
the shock pressure~\citep{sch09}, %
a scenario that implies fast shocks with efficient particle acceleration~\citep{blo01}.
The diffusion of cosmic rays in SNRs depends on the field orientation 
and the level of magnetic turbulence which, in turn, can result from the instability induced by the cosmic ray pressure gradient~\citep{dru12}.
MHD simulations
suggest that turbulent fields driven by hydrodynamic instabilities~\citep{ino13,ban16}  
 may have radially biased velocity dispersions, leading to selective amplification of the radial component and further cosmic ray production. 
 
The remnant of Supernova (SN) 1987A in the Large Magellanic Cloud 
has proven a unique laboratory to investigate particle acceleration in young SNRs \citep{zan17,pet17}. 
At the current stage of the evolution of the radio remnant, 
the synchrotron  emission observable at radio frequencies mostly originates from the shock wave interacting with 
high-density CSM in the equatorial plane (e.g. \citealt{zan13,zan14}, and references therein), distributed in a ring-like structure (equatorial ring, ER).
The emission around the ER can be fitted in the Fourier space via a thick torus 
~\citep{ng08,ng13},  although the shocks are now expanding above and below the equatorial plane  and interacting with high-latitude material confined within the nebula hourglass structure~\citep{pot14}.

This paper presents the results of polarimetric observations of SNR 1987A carried out with the Australia Telescope Compact Array (ATCA) between October 2015 and May 2016,  
from 20 
to 50 GHz. 
Details of the observations and data reduction  are given in \S~\ref{obs}. In \S~\ref{polintro}, we describe the extraction of the polarization components and introduce ad-hoc polarization parameters. The implications of the detection of polarized emission for the magnetic field in the SNR and, thus, particle acceleration and cosmic-ray production by the shock front, are discussed in \S~\ref{disc}.
\vspace{3mm}

\section{Observations} 
\label{obs}

SNR 1987A was observed at 7 and 15 mm wavelengths ($\lambda\lambda$) with the ATCA in 2015 October 16$-$19 and 2016 May 17$-$18, with $2\times12$ hr sessions on each frequency band in October 2015 and one 12-hr session on each band in May 2016.
In all sessions the ATCA was in the 6A configuration with a maximum baseline of 5939 m.
The observations were performed over $2\times2$ GHz bandwidth in each frequency band, centered on 22.2 and 23.7 GHz, and 43.4 and 49.0 GHz, respectively. 
Atmospheric conditions were optimal during all October sessions with rms of the path length fluctuations below 
300 $\mu$m.
In all observations, the standard bandpass calibrator PKS B0637$-$752 was observed for 2 minutes every 90 minutes while the phase calibrator PKS 0530$-$727 was observed for 1.5 minutes every 6 minutes on the source. 
Uranus was used as the flux density calibrator 
at 43.4 and 49.0 GHz. 
At 22.2 and 23.7 GHz, we used PKS B1934$-$638 as the primary flux density calibrator.
The data were reduced with {\sc miriad}\footnote[1]{http://www.atnf.csiro.au/computing/software/miriad/}.
Polarization leakage corrections were applied via the task  
\texttt{gpcal}, 
based on the polarization properties of the calibrator.
To avoid bandwidth depolarization, the data were reduced separately for 400-MHz sub-bands.
A weighting parameter~\citep{bri95} of robust $= 0.5$  was used in all bands for Stokes-$I$ images, and robust $=2.0$ was used for the derivation of Stokes-$Q$, $U$, and $V$ images. 
For Stokes$-I$ data, deconvolution was carried out via the maximum entropy method~\citep{gul78}, while no further deconvolution was performed on the Stokes-$Q$, $U$, and $V$ maps.
The integrated flux density of the Stokes-$I$ images is $\sim92$ mJy at 22 GHz and $\sim59$ mJy at 44 GHz.
The angular resolution, 
defined as the full width at half-maximum (FWHM) of the approximately gaussian central lobe of the restoring beam,
is $0\farcs4$ for the 22-GHz image and $0\farcs2$ for the 44-GHz map (Fig.~\ref{fig_22-44}).
\\

\section{Polarization measurements}
\label{polintro}

\subsection{Rotation measure}  
\label{equip}

For linearly polarized radio emission at short wavelengths 
 or Faraday-thin  objects~\citep{sok98}, 
the observed polarization angle $\psi$ is linked to the Faraday rotation measure ($\rm{RM}$) as~\citep{bur66}
\begin{equation}
	\psi = \psi_{0} + {\rm RM}\, \lambda^{2}, 
 	 \label{eq:RM1}
\end{equation}
with $\psi_{0}$ the intrinsic polarization angle.
For  wavelengths so close that
$\lvert{\lambda_{_{i+1}}^{2}-{\lambda_{_i}}^{2}}\rvert <\pi/2\, \rm{RM}_{0}$~\citep{ruz79},
the general definition of ${\rm RM} \approx{\rm d}\psi(\lambda)/{\rm d}\lambda^{2}$ 
can be taken as a linear function of $\lambda^{2}$, i.e.  
${\rm RM_{_{i+1,i}}} \approx (\psi_{_{i+1}}-\psi_{_{i}}) / (\lambda_{_{i+1}}^{2}-\lambda_{_{i}}^{2})$, being 
$\lambda_{_{i+1}}> \lambda_{_{i}}$.

\noindent
In our analysis, $\lambda_{_i}$ and $\lambda_{_{i+1}}$  
are taken as the central wavelengths of  
adjacent 400-MHz-wide frequency sub-bands, 
i.e. 
$\lambda_{_i}=\lambda_{0} +i\,\Delta\lambda$,    
and $i=0,1-3$ within each 2-GHz bandwidth (see \S~\ref{obs}).  
Due to high $\psi$ uncertainty associated with the fainter polarized emission at higher frequencies, the RM could not be estimated for sub-bands in the $\sim48.5-50.0$ GHz frequency range 
(Fig.~\ref{fig_PA-RM}).

\vspace{2mm}
\subsection{Linear polarization components} 
\label{components}

The linear polarized intensity, $I_{_{P}}=\sqrt{Q^{2}+U^{2}}$, 
is the modulus of the Q and U Stokes parameters.
The $I_{_{P}}$ distribution across SNR 1987A as observed at 22 GHz
is shown in Fig.~\ref{fig_00_pol}, blanked for polarized emission intensity lower than $2\sigma$.
We note that bias in the observed linear polarization 
arises when the measurement of $Q$ and $U$ is significantly affected by noise, as 
the quantity $Q^2+U^2$ is overestimated~\citep{sim85}. 
Assuming the errors on the actual Stokes parameters, $Q_{0}$ and $U_{0}$, are known and 
both equal to $\sigma$, 
i.e. $Q=Q_{0}\pm\sigma$ and $U=U_{0}\pm\sigma$,  
the true degree of polarization can be taken as~\citep{sim85} $I_{_{P0}}=\sqrt{Q_{0}^2+U_{0}^2}$ 
if $I_{_{P0}}/\sigma>4$, 
 with polarization angle $\psi_{0}=1/2\,\, {\rm tan}^{-1}\,Q_{0}/U_{0}$. 
In the case of our $3\sigma$ polarization measurements,  $I_{_{P}}$ is 
overestimated by $\sim2\%$ at 22 GHz with an error on  
$\psi_{_{0}}$ of less than $5^{\circ}$ 
over the ring-like structure of the SNR, while at 44 GHz 
$(I_{_{P}}-I_{_{P0}})/I_{_{P0}}\approx\ 3\%$ 
and 
$\psi=\psi_{_{0}}\pm7^{\circ}$ 
over the brightest Stokes-$I$ sites.

Similarly, the standard expression of the fractional polarization as a function of the Stokes parameters, 
$P\equiv\sqrt{Q^{2}+U^{2}}/I$,  
is greatly affected by bias
when the signal-to-noise ratio is low.
To bypass this bias, 
we introduce two polarization parameters, quasi-$E$ (or $\widehat{E}$) and quasi-$B$ (or $\widehat{B}$) polarizations, which can be derived for emission distributions characterized by polar axis symmetry.
These quasi-polarizations 
are obtained via polar transformation of the 
linear polarization Stokes parameters, $Q$ and $U$, i.e.  
\begin{equation}
\begin{cases}
	\widehat{E}=U\,\textrm{sin}(2\chi)+Q\,\textrm{cos}(2\chi) \\
	\widehat{B}=U\,\textrm{cos}(2\chi)-Q\,\textrm{sin}(2\chi), 
\end{cases}
 \label{eq:BE}
\end{equation}
where 
$\chi$ is  the position angle 
of the $Q$ and $U$ 
measurements relative to the 
central reference.
For radio sources with polar morphology such as SNRs, 
the $\widehat{E}$ and $\widehat{B}$ parameters 
are the orthogonal components of the electric 
field.  
In the specific case of SNR 1987A,
the central reference is taken as the supernova site~\citep{rey95},
and the position angle $\chi$ is measured from north to east (see Fig.~\ref{fig:EB_images}).
After correction for Faraday rotation, 
the $\widehat{E}$ and $\widehat{B}$ quantities
trace the 
components of the magnetic field, aligned at 
$\left[0^{\circ}, 90^{\circ}\right]$  and $\left[-45^{\circ}, +45^{\circ}\right]$ to the tangent of the ring-like structure of the SNR, 
analogous to the $E$ and $B$ polarization modes~\citep{zal97}. 
As for the $E$ sign convention, negative values of $\widehat{E}$ identify a radial pattern of the magnetic field, which we signify as $\boldsymbol{B_{\parallel}}$.  
Fitting of the $\widehat{E}$ vs $I$  distribution 
allows a robust estimate of the  fractional polarization in regions of varying brightness and, especially, in low-emissivity sites.

\vspace{5mm}
\section{Discussion} 
\label{disc}

\begin{figure*}[htb]
\centering
\vspace{-0mm}
\advance\leftskip-9mm
\includegraphics[angle=0, width=193mm]{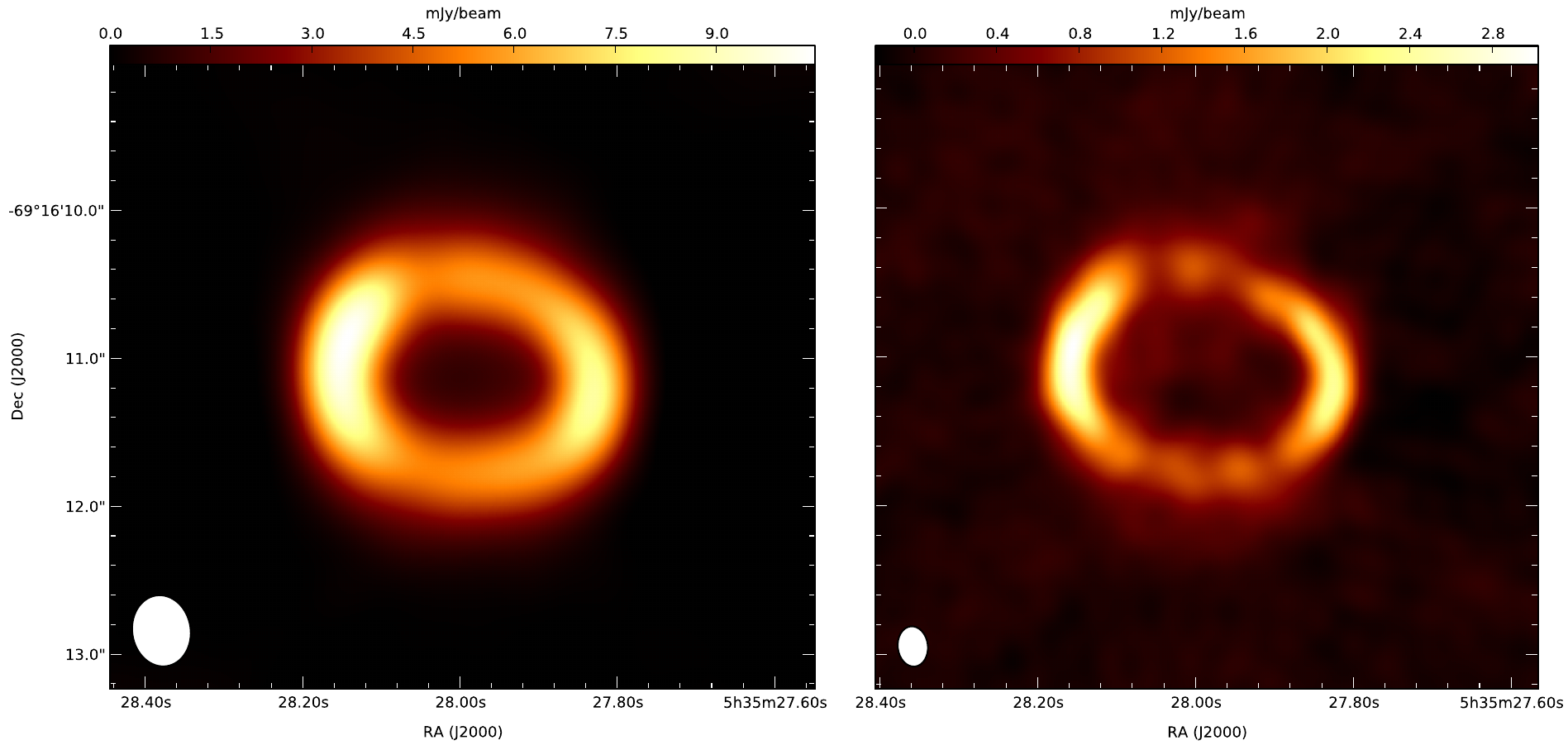}
{\caption{
\noindent
Diffraction-limited Stokes-$I$ continuum images of SNR 1987A at 22 GHz (left) and 44 GHz (right) obtained from ATCA observations carried out from October 2015 to May 2016.
The beam size (FWHM) is $0\farcs48\times0\farcs39$ at 22 GHz and $0\farcs27\times0\farcs20$ at 44GHz, as plotted in the lower left corner of each image.
 \label{fig_22-44}
 }}
\end{figure*}

\begin{figure*}
\vspace{-20mm}
\advance\leftskip-4.0mm
\includegraphics[angle=0, width=170mm]{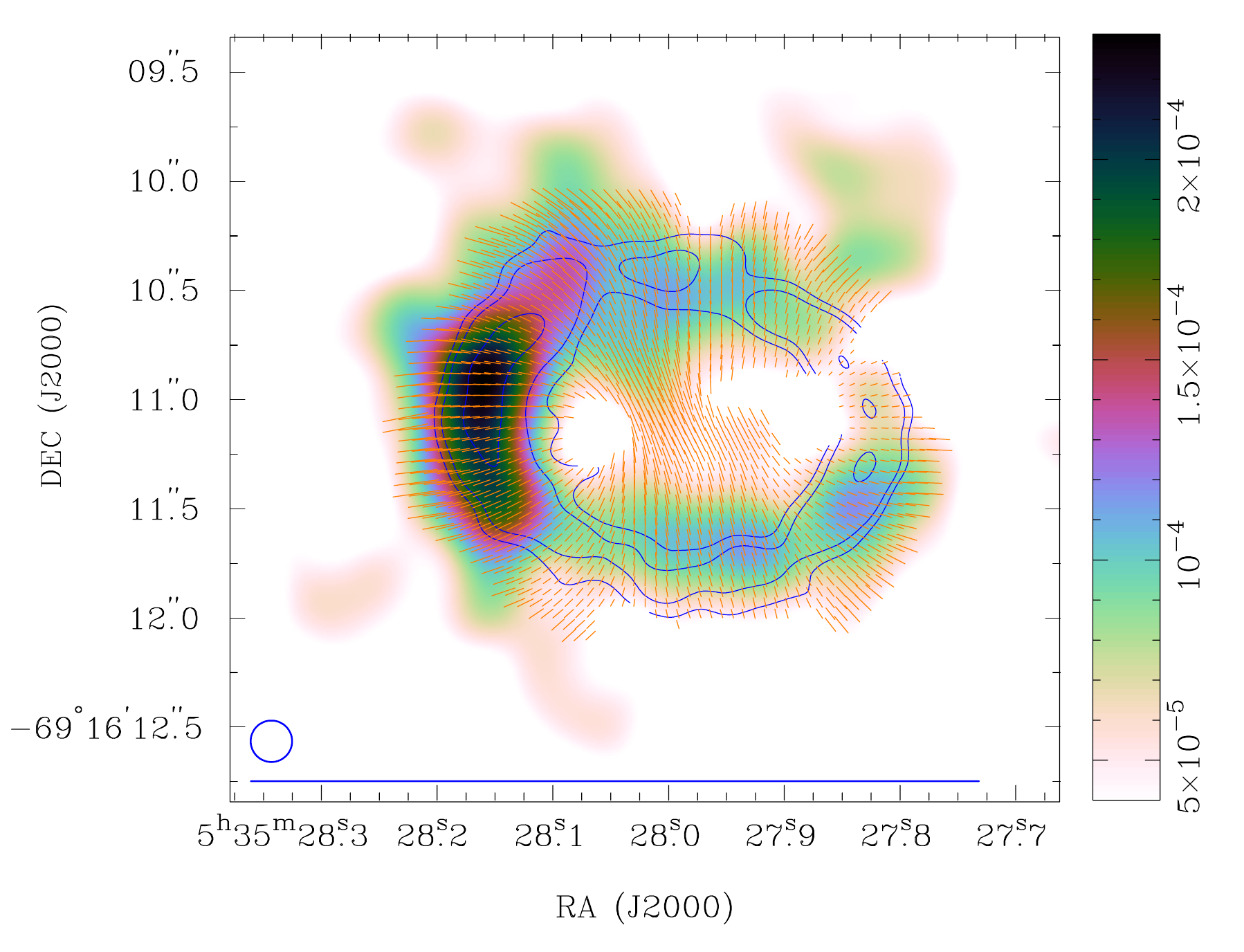}
\caption{
Map of the polarized intensity, defined as $I_{P}=\sqrt{Q^{2}+U^{2}}$, where $Q$ and $U$ are the linear polarization Stokes parameters, generated for observations of SNR 1987A at 22 GHz. The map has an angular resolution (FWHM) of $0\farcs4$ and is shown with colour scale in Jy beam$^{-1}$.
The map is overlaid with the contours (blue) of the  contemporaneous Stokes-$I$ image of the SNR at 44 GHz, 
restored with a $0\farcs2$ circular beam (bottom left corner). The 44 GHz contours are shown at flux density levels of  0.4, 6.0, and 1.5 mJy beam$^{-1}$. 
The polarization vectors have been rotated by 90$^\circ$ to show the intrinsic 
magnetic field orientation (orange-coloured lines). Vectors in the central region of the remnant are detections between $2\,\sigma_{P}$ and $3\,\sigma_{P}$, where $\sigma_{P}=22.3$ $\mu$Jy beam$^{-1}$ 
the mean standard deviation of the noise in the Stokes-$Q$ and $U$ images.
The vectors length 
 is proportional to the fractional polarization $P\equiv\sqrt{Q^{2}+U^{2}}/I$; the line extent corresponding to $P=100\%$ is 
shown at the bottom. 
 \label{fig_00_pol}
 }
\end{figure*}

\begin{figure*}[htb]
	\begin{minipage}[c]{180mm}
		\begin{center}
		\vspace{-20mm}
		\advance\leftskip+1mm
		\includegraphics[width=180mm, angle=0]{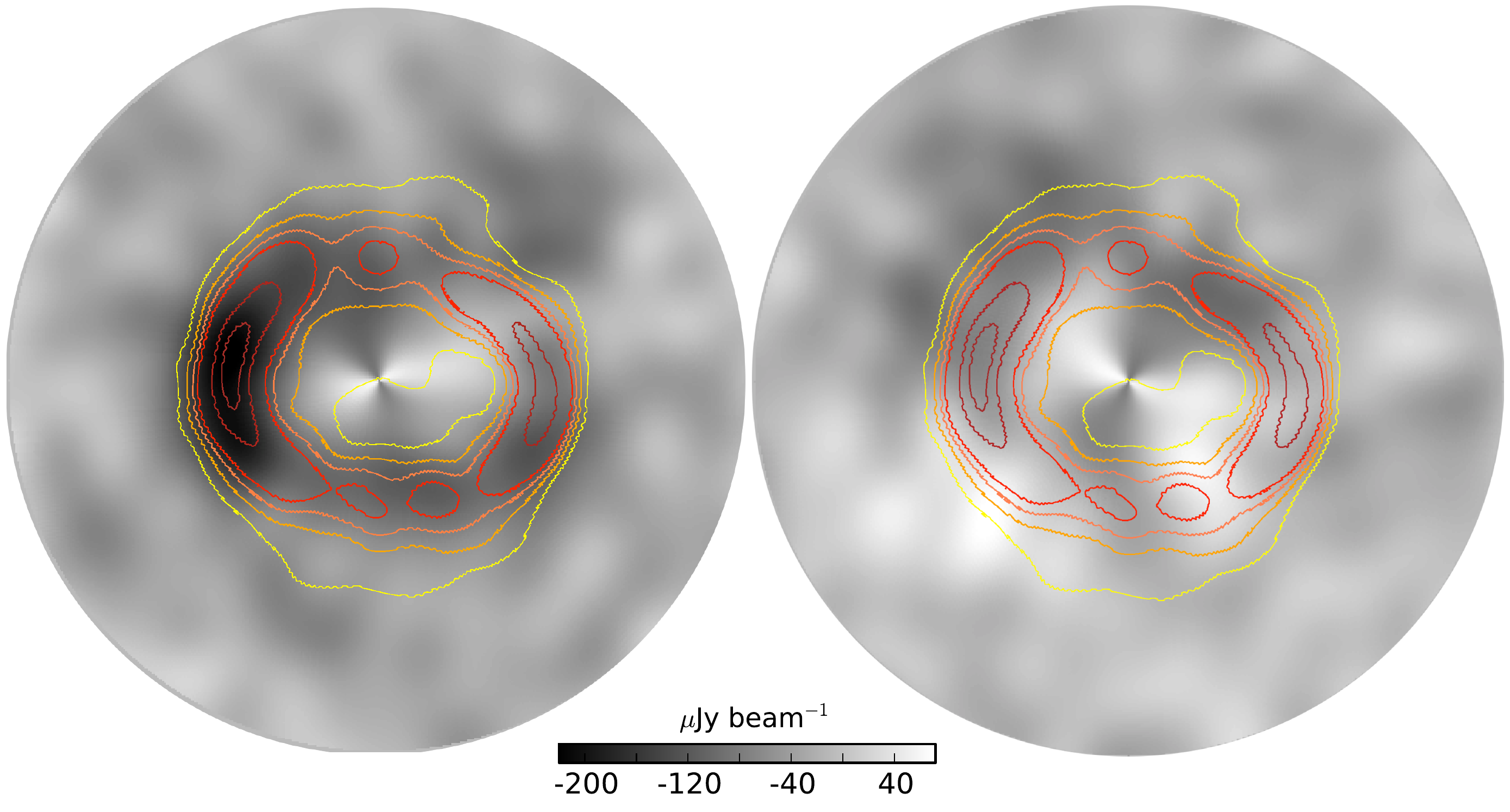}
		\end{center}
	\end{minipage}
	\begin{minipage}[c]{180mm}
		\begin{center}
		\vspace{-0.0mm}
		\advance\leftskip+2mm
		\includegraphics[trim=10mm 6.0mm 6.0mm 10mm, clip=true,width=185mm, angle=0]{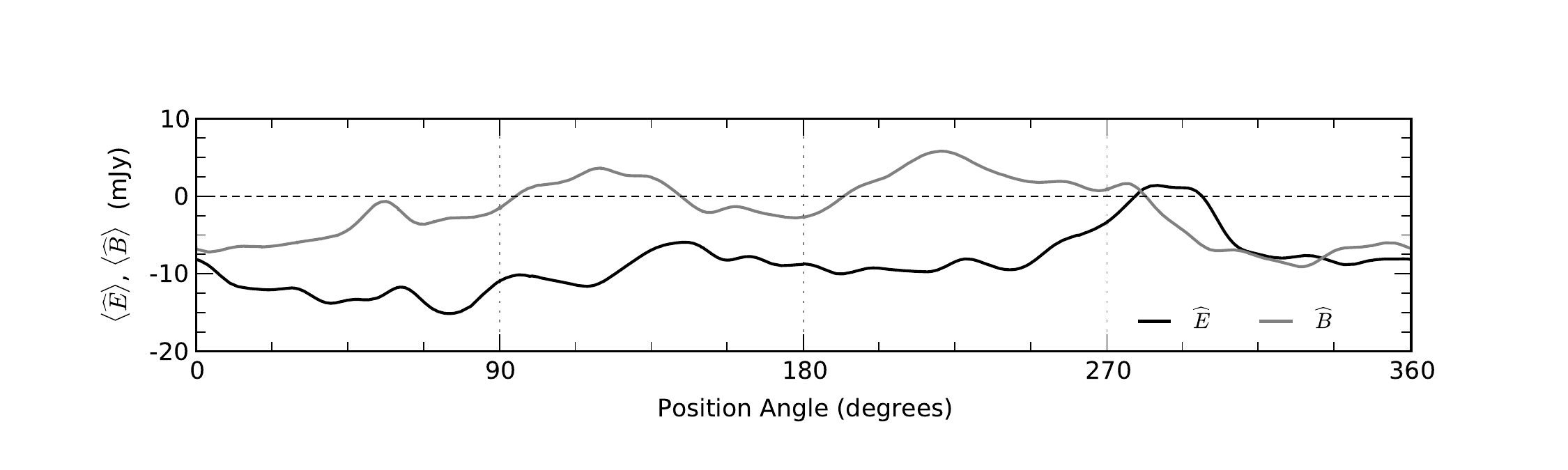}
		\end{center}
		\end{minipage}
\vspace{3.0mm}
\caption{
Maps of the $\widehat{E}$ (left) and $\widehat{B}$ (right) polarizations generated for observations of SNR 1987A at 22 GHz. 
From the linear polarization Stokes parameters $Q$ and $U$, 
$\widehat{E}=U\,\textrm{sin}(2\chi)+Q\,\textrm{cos}(2\chi)$
and
$\widehat{B}=U\,\textrm{cos}(2\chi)-Q\,\textrm{sin}(2\chi)$, where 
$\chi$ is  the position angle 
measured from north to east.
 The map derivation in polar coordinates is centered on the SN site  [RA $05^{\rm h}\;35^{\rm m}\;27\fs968$, Dec $-69^{\circ}\;16'\;11\farcs09$ (J2000)]~\citep{rey95}, 
and has a $2^{\prime\prime}$ radius. 
In this coordinate system, a negative $\widehat{E}$ is equivalent 
to a tangential polarization vector and therefore 
to the radial 
component of the magnetic field, $\boldsymbol{B_{\parallel}}$,  
i.e. parallel to the normal of the shock-front plane, while weaker non-radial field components can be traced by the  negative $\widehat{B}$. 
Both maps are superimposed with the contours of the Stokes-$I$ intensity map at 44 GHz, which has a resolution of $0\farcs2$ (FWHM). 
The contours are shown at 14\%, 22\%, 30\%, 38\%, 70\%, 
and 90\% flux density levels, with a colour scheme from yellow to brown to identify regions of increasing brightness. 
The integrated intensity of the $\widehat{E}$/$\widehat{B}$ parameters for each subtended angle (north to east, with north at PA $\equiv0^{\circ}$ and east at PA $\equiv90^{\circ}$) is shown in the bottom plot. The $\widehat{E}$ polarization  intensity is integrated along a $2^{\prime\prime}$ 
 radius and shown in black,  while the integrated intensity of the $\widehat{B}$ polarization is plotted in grey.
\label{fig:EB_images}} 
\end{figure*}

As it can be seen in the Stokes-$I$ flux density distribution at mm $\lambda\lambda$ (Fig.~\ref{fig_22-44}), the radio emission from the remnant of SN 1987A
currently  
extends beyond the ER, with linearly polarized emission in both the inner region of the SNR and over the ER (Fig.~\ref{fig_00_pol}).
The derivation of $\widehat{E}$ and $\widehat{B}$ via Eq.~\ref{eq:BE} 
(Fig.~\ref{fig:EB_images}) allows to 
map the marked radial component of the magnetic field 
($\boldsymbol{B_{\parallel}}$), which appears especially strong
 over the brightest regions on the eastern lobe 
 (see $\widehat{E}$ map in Fig.~\ref{fig:EB_images}). 
The radial alignment  
is observed to be maintained through the ER,
thus encompassing both the reverse and forward shocks, 
i.e. where  the majority of the synchrotron emitting electrons is
generated~\citep{pot14}.
While a predominantly 
radial magnetic field
has been found in many 
young SNRs~\citep{mil87}, SNR 1987A is by far the youngest  remnant to exhibit such alignment.
Weaker non-radial 
field components
appear localised in the NE sector just outside the brightest regions of the remnant, as well as in the faintest Stokes-$I$ sites on the ring, %
i.e. at  position angle PA$\,\sim225^{\circ}$ and PA$\,\sim325^{\circ}$ 
(Fig.~\ref{fig:EB_images}).

Since the synchrotron brightness directly tracks the magnetic field strength, 
the association of the strongest  $\boldsymbol{B_{\parallel}}$ field with the high-emissivity sites on the eastern lobe of the SNR is consistent with the 
scenario of large injection efficiency and 
amplification of the magnetic field due to cosmic ray production.
The expansion velocity 
extrapolated for the faster eastbound shocks~\citep{zan13}, $u\sim6000$ km s$^{-1}$, 
would yield high upstream Alfv\'enic Mach numbers, $M_{A}$, depending on the shock arrangement,
being $M_{A}=u/v_{A}$ with $v_{A}\ll u$ the speed of the Alfv\'en waves generated by the cosmic rays.  
The  detection 
 in Saturn's strong  bow shock of electron acceleration under 
quasi-parallel magnetic conditions~\citep{mas13}, 
suggests that when $M_{A}\sim100$ quasi-parallel shocks become very 
effective electron accelerators. 
Globally high
Mach numbers linked to quasi-parallel shocks 
likely result from non-linear amplification of the magnetic field due to very efficient cosmic-ray acceleration~\citep{bel01},   
being
$M_{A}\propto(B_{_{\rm SNR}}/B_{0})^{2}$,
where $B_{_{\rm SNR}}$ is the magnetic field strength within the SNR and
$B_{0}$ is the magnetic field near the remnant.

The magnetic field strength within the SNR
can be inferred from the energy equipartition and pressure equilibrium between the remnant magnetic field and cosmic rays \citep{bec05,arb12}, which is
\vspace{-2mm}
\begin{equation}
	B_{_{\rm SNR}} \approx \biggl[ G_{0} \,G \,(\mathcal{K}+1) \, { S_{\nu} \over f d\, \theta_{_{\rm SNR}}^{3}} \,\,
	\nu^{\scalebox{1.1}{$ {(1-\gamma)\over2} $} }  \biggr]^{\scalebox{1.1}{$ { 2 \over (5+\gamma)} $}},
	\label{eq:Beq}
\end{equation}
where $G_{0}$ is a constant, $G=G(\nu,\gamma)$ is the product of different functions varying with the minimum and maximum frequencies associated with the spectral component and the synchrotron spectral index~\citep{bec05},  
$\mathcal{K}$ is the ion/electron energy ratio, 
$f$ is the volume filling factor of radio emission, 
$\theta_{_{\rm SNR}}=R_{_{\rm SNR}}/d$ is the angular radius,  and
$\gamma=1-2\alpha$ with $\alpha$ the synchrotron spectral index,
being the synchrotron emission $S_{\nu}\propto\nu^{\alpha}$.
Considering $1\farcs0\lesssim R_{_{\rm SNR}}\lesssim1\farcs1$, 
 $-0.95\le\alpha\le-0.91$,  
$20\le\nu\le50$ GHz, 
$S_{\nu}= 92$ mJy at $\nu=22$ GHz, and taking $\mathcal{K}\approx100$, 
while $f\approx0.5$,  
Eq.~\ref{eq:Beq} yields 
 $B_{_{\rm SNR}}\sim 2$ mG. 
We note that although the equipartition of the magnetic field is a conjecture for young SNRs, 
it is considered applicable~\citep{arb12,sok98} for remnants or specific SNR sites where  $-1.0\lesssim\alpha\lesssim-0.8$, or for energy spectral indices  $2<\gamma\lesssim3$. 
These conditions are met
in the brightest eastern sites of SNR 1987A, i.e. in the regions that have been consistently associated with steeper synchrotron spectral indices (\citealp{zan14,zan13}). 

\begin{figure*}
\centering
\includegraphics[angle=0, width=16.5cm]{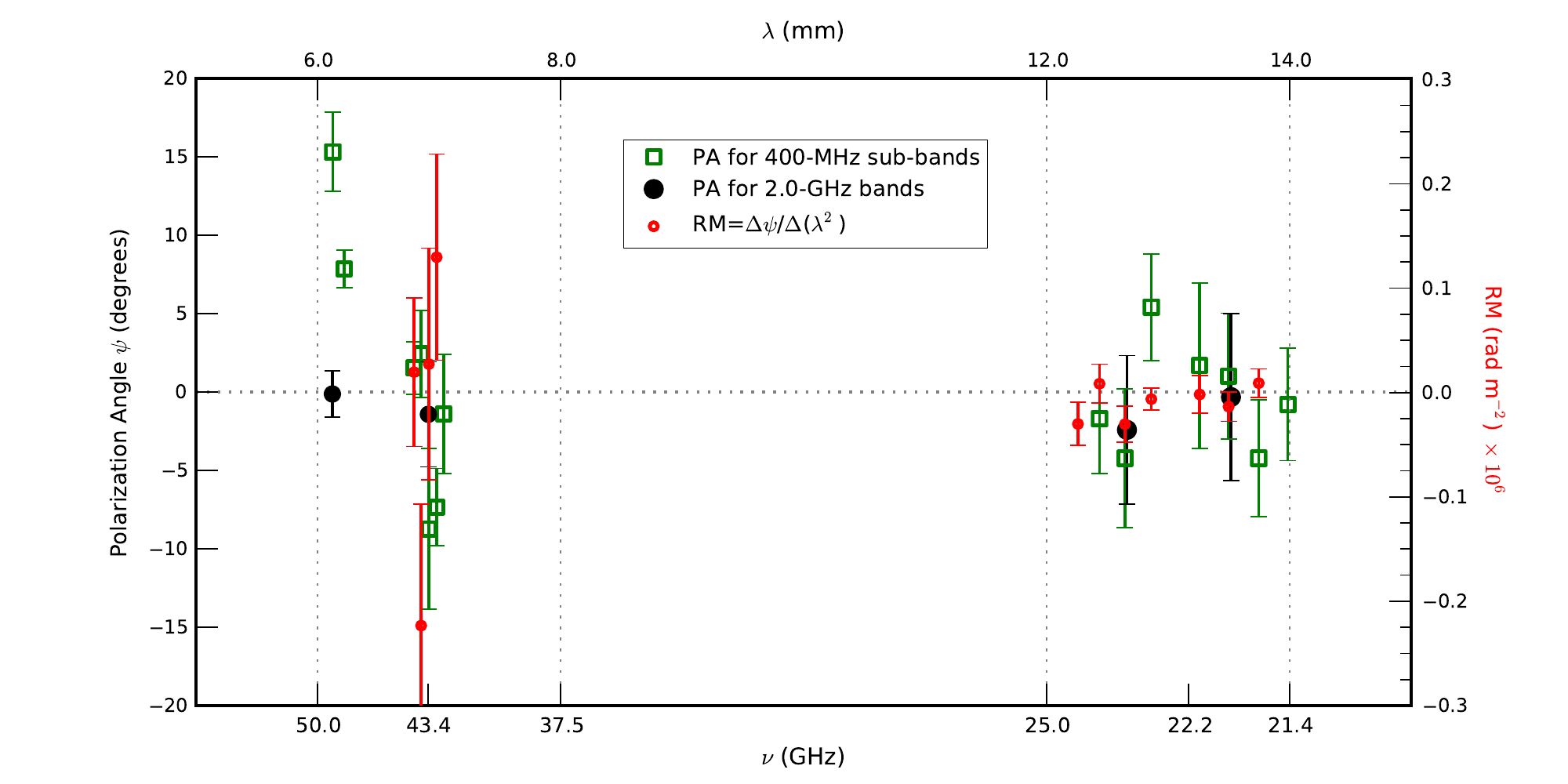}
{\caption{
\noindent
Polarization angle ($\psi$) and rotation measure ($\textrm{RM}$) vs frequency.
The wavelength dependence of 
$\psi$, 
being $\textrm{tan}(2\psi)=U/Q$, 
and of RM,  
given $\textrm{RM}=\Delta\psi/\Delta(\lambda^{2})$,
is investigated for all observations of polarized emission from 20 to 50 GHz, from October 2015 to May 2016, with reference to the compact brightest region on the eastern lobe (see Fig.~\ref{fig_22-44}). 
\label{fig_PA-RM}
}}
\end{figure*}

We determine the strength of the ambient magnetic field, $B_{0}$, i.e. the magnetic field within the CSM near the SN, from the RM (Fig.~\ref{fig_PA-RM}). 
The magnetic field along the line of sight ($\textrm{los}$) can be linked to the RM as
\begin{equation}
\mathrm{RM}\approx \frac{ e^{3} \, \lambda^{2}}{2\pi(m_{\mathrm{e}}c^{2})^{2}}\int_{\mathrm{los}}n_{\mathrm{e}}(l)B_{\rm{los}\parallel}(l)\,{\rm d}l.
 \label{eq:Blos}
\end{equation}

\vspace{2mm}
\noindent
Given $\rm{RM}\approx1.3\times10^5$ rad m$^{-2}$, as the upper limit at $\sim7$ mm wavelength 
(see Fig.~\ref{fig_PA-RM}),
and considering that  the medium in which the SN radio emission propagates has an electron density $n_{e}\sim110$ cm$^{-3}$,
as from low-frequency measurements by \citet{cal16}, 
Eq.~\ref{eq:Blos} yields $B_{\rm{los}\parallel}\simeq B_{0} \approx 28$ 
$\mu$G.

In this context, 
a non-linear  magnetic field amplification, $(B_{_{\rm SNR}}/B_{0})^\theta$ with $1<\theta\leq 2$, 
would lead to 
$100 \lesssim M_{A}\lesssim 10^{3}$.
For such Mach numbers, strong fluctuations in $B$-field strength would invoke  
non-linear cosmic-ray-excited turbulence~\citep{bel04}.

\begin{figure*}[htb]
\begin{minipage}[c]{260mm}	
	\begin{center}
		\begin{minipage}[c]{125mm}
			\begin{center}
			\vspace{-117mm}
			\advance\leftskip30mm    											
			\includegraphics[width=118mm, angle=0]{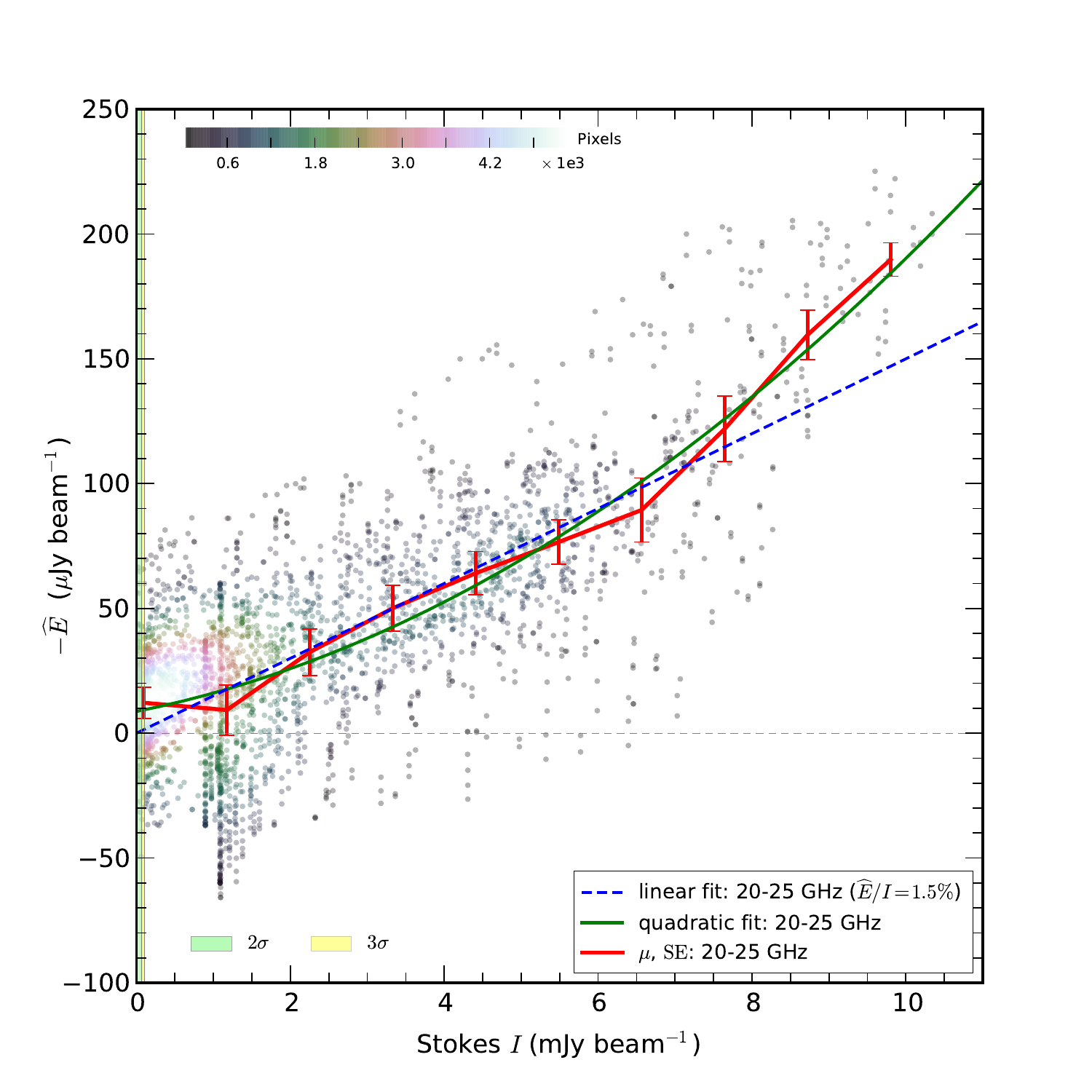}
			\end{center}
		\end{minipage}
		\begin{minipage}[c]{125mm}
			\begin{center}
			\vspace{95.0mm}
			\advance\leftskip-200.6mm
			\includegraphics[width=118mm, angle=0]{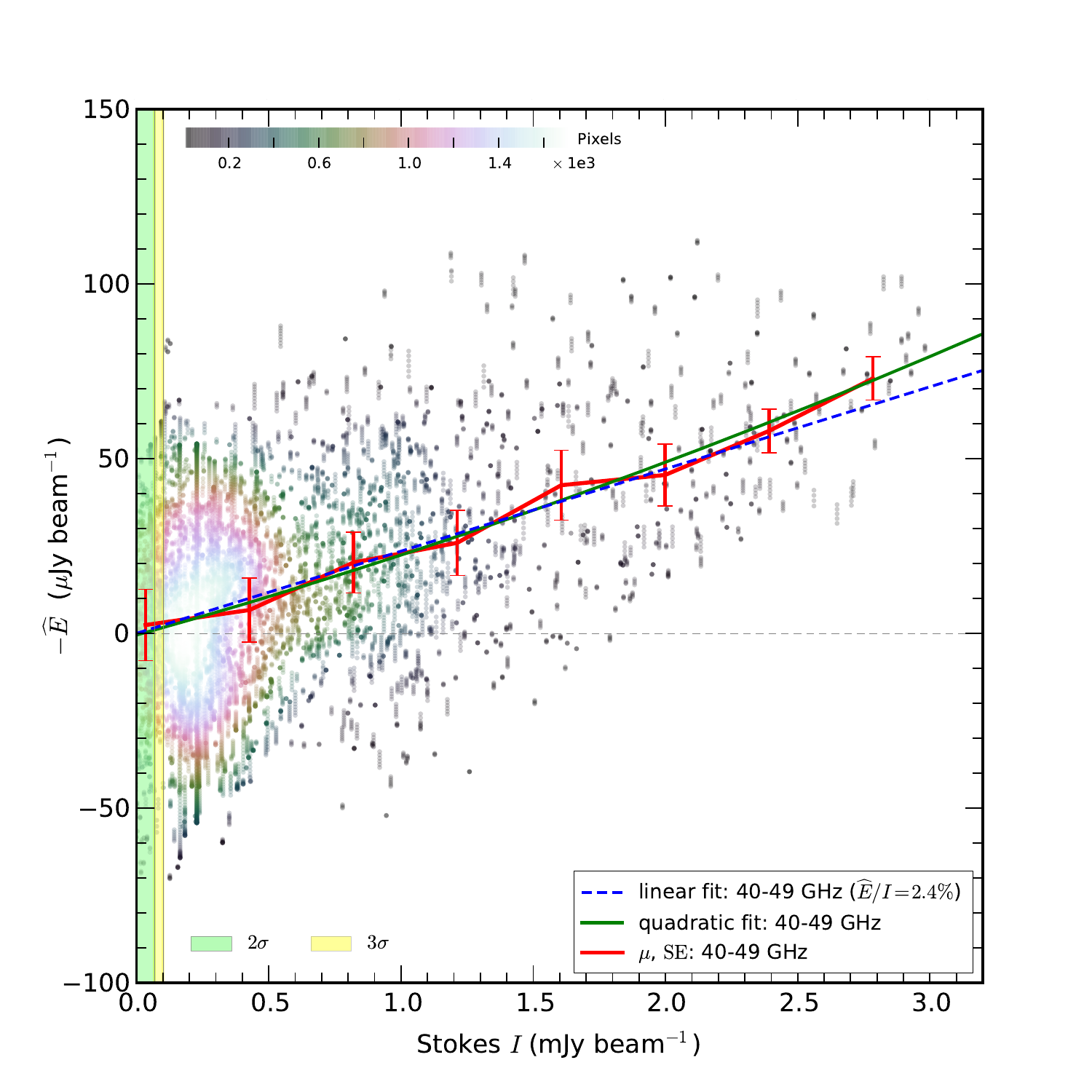}
			\end{center}
		\end{minipage}
	\end{center}
\end{minipage}
\caption{
Polarized vs unpolarized intensities at 22 GHz (top) and 44 GHz (bottom) via $\widehat{E}-I$ plots.
The  Stokes-$I$ and $\widehat{E}$ images at 22 and 44 GHz  
are binned in $0\farcs1\times0\farcs1$ pixels, with density distribution as shown in the colorbar in the top left corner. 
Since the angular resolution (FWHM) of the images is $0\farcs4$ at 22 GHz and $0\farcs2$ at 44 GHz, the image pixels have been sampled to limit the correlation within the restoring beam.
The $\widehat{E}-I$ data have been further binned to derive the mean ($\mu$) and the standard error (SE). 
The light green and yellow rectangular regions highlight the 2$\sigma$ and 3$\sigma$ thresholds of the 
Stokes-$I$ image. Linear and quadratic fits are plotted with dashed blue and green lines, respectively.
\label{fig:EI_plots}} 
\vspace{-5mm}
\end{figure*}

The relative ratio of the coherent and disordered magnetic field components
can be assessed via inspection of the degree of polarization of the synchrotron emission.
As introduced in \S~\ref{components}, we use $\widehat{E}-I$ plots to determine the fractional polarization in the SNR.
The $\widehat{E}-I$ plots derived for observations at 22 and 44 GHz  (Fig.~\ref{fig:EI_plots}) show that the degree of polarization is mostly constant across the remnant. %
From linear fitting of the $\widehat{E}-I$ distribution (dashed blue lines in Fig.~\ref{fig:EI_plots}),  the overall degree of polarization 
is $1.5\pm0.2\%$ at 22 GHz and $2.4\pm0.7\%$ at 44 GHz. 
The mean fraction of polarized emission 
 in the brightest sites 
 on the eastern lobe 
 is $2.7\pm0.3\%$  at  22 GHz and $3.5\pm0.7\%$ at 44 GHz. 
For comparison, observations of Cassiopeia A (Cas A, SN $\sim$1680) at 19 GHz yielded 4.5\% of linear  polarization  around  the rim  and  absence of polarized emission in the centre~\citep{may68}. 
Analysis of the $P$ distribution against the total 
intensity in Cas A has not 
revealed any significant correlation~\citep{and95}, while sites of very faint synchrotron emission in the remnant  of SN 1006 have been associated with $P$ values close to the theoretical maximum~\citep{rey13}.
The $\widehat{E}-I$ plots hint at an increased fraction of polarized emission in the high-emissivity sites of SNR 1987A. This trend is more marked in the higher signal-to-noise observations at 22 GHz,  where the fractional polarization appears to increase super-linearly as the emission becomes brighter, and for $I \gtrsim6.5$ mJy beam$^{-1}$ the $\widehat{E}-I$ distribution is better described by a quadratic fit (green line in Fig.~\ref{fig:EI_plots}). 
While a relatively low degree of polarization is not an unequivocal
indicator of the extent of the ordered component of the magnetic field,  
a direct correlation between increasing fractional polarization and brighter emission sites would require high efficiency rates of cosmic-ray production, possibly achieved by  short-scale turbulent amplification of the magnetic field interacting with the 
dense clumps of the CSM~\citep{mei14}. 
We note that the beam depolarization associated with our observations hampers an accurate estimate of the local degree of polarization, especially if  the magnetic field undergoes micro-instabilities by the shock front and the downstream regions~\citep{mar16}.

As regards the central region of SNR 1987A,  
from the distribution of the $\widehat{E}$ and  $\widehat{B}$ polarizations 
shown in Fig.~\ref{fig:EB_images}, 
at 22 GHz the inner magnetic field 
appears to have a 
prevalence of non-radial components along the NW-SE axis, which extend to the outer edge of the  ring.
The derivation of $P=P(Q,U,I)$ for the inner remnant 
yields $P=3.6\pm1.5\%$ at 22 GHz. 
Although the detection of polarized emission 
 flags the presence of magnetized shocks in the centre of the remnant, the $2\sigma$ detection can not be used for a meaningful estimate of the fractional polarization,
as the beam depolarization is rather significant for the expected size of a possible pulsar wind nebula~\citep{zan14}. 

\vspace{5mm}
GZ acknowledges the support of the International Centre for Radio Astronomy  Research (ICRAR).
The Australia Telescope Compact Array is part of the Australia Telescope National Facility, which is funded by the Australian Government for operation as a National Facility managed by CSIRO.
\vspace{5mm}

\invisiblesection{References}

\bibliographystyle{apj}

%
%

\pagebreak

%
%

%
%

\pagebreak

%
%

\pagebreak

%
%

\pagebreak

\clearpage

\end{document}